\documentclass[english,prb,twocolumn,superscriptaddress]{revtex4-1}
\usepackage{graphicx}
\usepackage{amsmath}
\usepackage{babel}
\begin{document}

\title{Determinstic Generation of a Quantum Dot-Confined Triexciton and
its Radiative Decay via Three-Photon Cascade }

\author{E. R. Schmidgall}
\affiliation{Department of Physics and the Solid State Institute,
The Technion - Israel Institute of Technology, Haifa 32000, Israel}
\author{I. Schwartz}
\affiliation{Department of Physics and the Solid State Institute,
The Technion - Israel Institute of Technology, Haifa 32000, Israel}
\author{L. Gantz}
\affiliation{Department of Physics and the Solid State Institute,
The Technion - Israel Institute of Technology, Haifa 32000, Israel}
\affiliation{Department of Electrical Engineering, The Technion - Israel Institute of Technology,
Haifa 32000, Israel}
\author{D. Cogan}
\affiliation{Department of Physics and the Solid State Institute,
The Technion - Israel Institute of Technology, Haifa 32000, Israel}
\author{S. Raindel}
\affiliation{Department of Electrical Engineering, The Technion - Israel Institute of Technology,
Haifa 32000, Israel}
\author{D. Gershoni}
\affiliation{Department of Physics and the Solid State Institute,
The Technion - Israel Institute of Technology, Haifa 32000, Israel}
\email{dg@physics.technion.ac.il}

\begin{abstract}
Semiconductor quantum dots (QDs) have potential applications in quantum
information processing due to the fact that they are potential on-demand sources
of single and entangled photons. Generation of
polarization-entangled photon pairs was demonstrated using the biexciton-exciton radiative cascade.
One obvious way to increase the number of quantum correlated
photons that the QDs emit is to use higher-order multiexcitons,
in particular the triexciton. Towards achieving this goal, we first demonstrate
deterministic generation of the QD-confined triexciton in a well-defined coherent state and then spectrally identify and directly measure a three-photon radiative cascade resulting from the sequential triexciton-biexciton-exciton radiative recombination.
\end{abstract}


\maketitle
Semiconductor quantum dots (QDs) confine charge carriers in three
spatial dimensions. This confinement results in a discrete spectrum
of energy levels and energetically sharp optical transitions between
these levels. These ``atomic-like'' features, together with their
compatibility with semiconductor-based microelectronics and optoelectronics
make QDs promising building blocks for future technologies involving
quantum information processing (QIP) \citep{imamoglu1999,loss1998} which rely on single photon detectors, single photon emitters and light-matter interaction involving single photons and single carriers.
Devices that emit single and entangled photons on demand are typical
examples of these potential applications.\citep{Santori,yuan2002,akopian2006,sellinart2010,muller2014}
QDs are known sources of polarization-entangled photon pairs resulting from the biexciton-exciton
radiative cascade.\citep{akopian2006,young2006,hafenbrak2007} Using pulsed laser excitation,
this entangled photon pair emission process can, in principle, be performed on-demand.\citep{muller2014}
Deterministic generation of higher-order multiexcitons is a conceptual way for increasing the number of quantum correlated photons that a single QD emits. Here, we study the neutral triexciton ($XXX^{0}$) which contains three electron-hole (e-h) pairs, as a candidate for achieving this goal.

First, we discuss the optical transitions from the QD-confined
ground neutral triexciton states to various biexciton ($XX^{0}$) states. We then experimentally identify these optical transitions using photoluminescence (PL) and PL excitation spectroscopy.
We use this information to demonstrate deterministic triexciton generation using a sequence of three non-degenerate laser pulses. We conclude by using third-order intensity cross-correlation
measurements to demonstrate a triexciton-biexciton-exciton radiative cascade for the first time.

While there are previous reports on radiative cascades from triexcitons in single QDs,\citep{arashida2011,persson2004} these reports fell short of identifying the triexciton fine structure, let alone demonstrating its deterministic generation or a three photon radiative cascade.

\begin{figure}
\includegraphics[width=1\columnwidth]{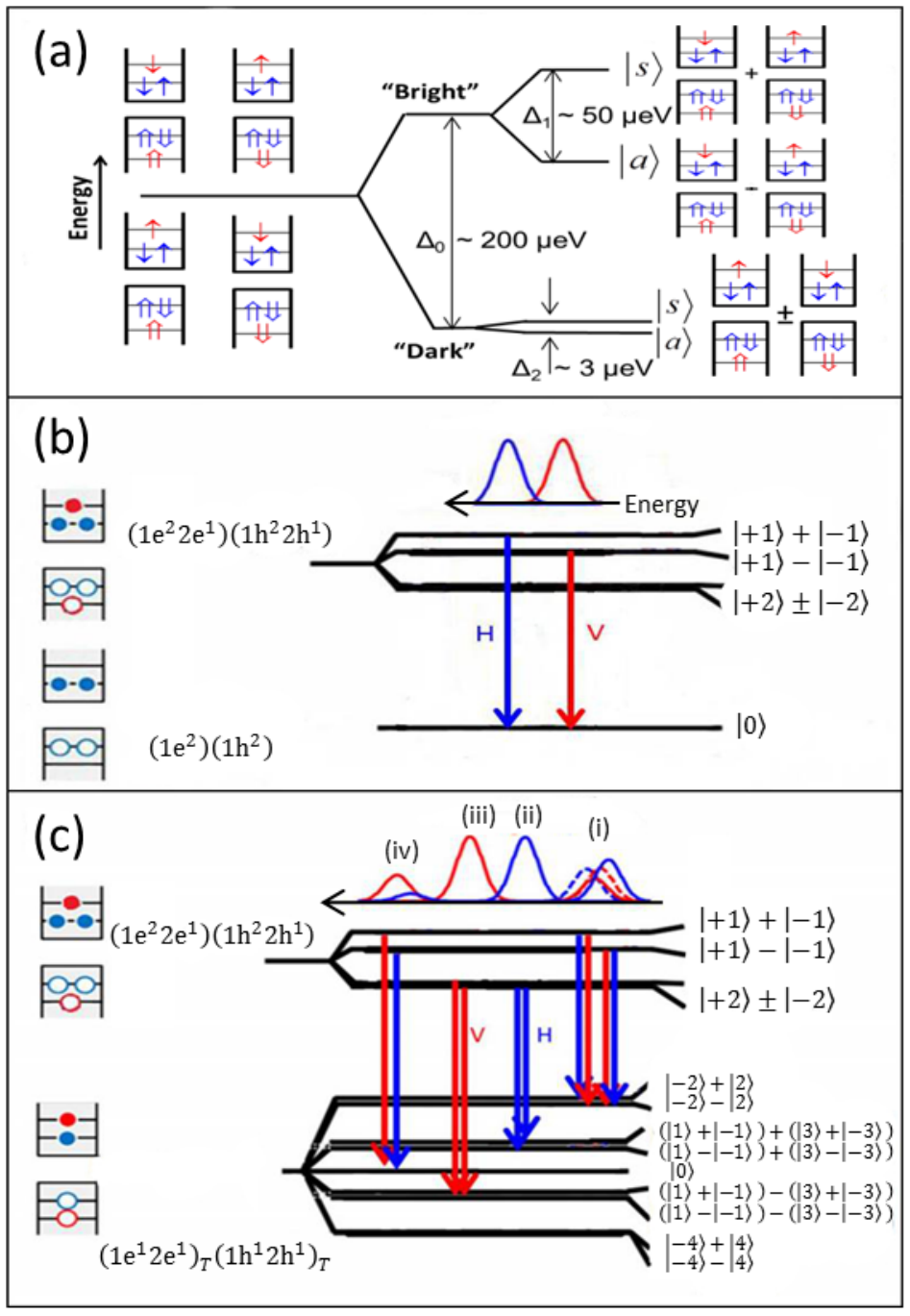}
\protect\caption{(a) Schematic illustration of the QD-confined ground-state triexciton.
$\uparrow(\Uparrow)$ represents an electron (heavy hole). The spin
projection on the growth direction corresponds to the arrow direction. 
 (b) [(c)] Allowed optical transitions
from the triexciton states to the ground [excited e-triplet/h-triplet] biexciton state[s]
Horizontal, H, (vertical, V) rectilinearly polarized emission
is indicated by blue (red) arrows. The total spin projection of the
initial (final) state is presented to the right of the energy level.
The calculated~\citep{benny2011prb} PL emission spectrum (above the arrows) was obtained from a many-carrier model including direct and exchange Coulomb interactions between confined carriers. Allowed transitions and intensities were calculated using the dipole approximation.\cite{poem2007}
\label{fig:theory}}
\end{figure}

The QD confined $XXX^{0}$ contains three electron-hole (e-h) pairs. The population of these carriers in their ground state can be approximately described as a pair of electrons and a pair of heavy holes occupying their respective ground energy level forming an antisymmetric spin configuration (a singlet). In addition, there is one unpaired electron and heavy hole in their respective second energy level. It follows, therefore, that the triexciton fine structure fully resembles that of the exciton ($X^{0})$ ~\citep{bayer2002,takagahara2000,ivchenko,poem2007} where the unpaired e-h exchange interaction fully removes the degeneracy between the four possible e-h spin configurations, as schematically described in Fig. \ref{fig:theory}(a)  There are two ``bright" triexciton states in which the e-h pair spins are antiparallel and two ``dark" triexciton states in which the e-h pair spins are parallel. Similarly to the exciton case, the degeneracy between the ``dark" and ``bright" triexciton pairs is further removed by the anisotropic and short range e-h exchange interactions as seen in Fig. \ref{fig:theory}(a).

Radiative recombination between the triexciton e-h pairs occurs only between pairs of antiparallel spin directions.~\cite{bastard,pawel} In addition to opposite spins, efficient recombination requires also significant e-h spatial envelope wavefunction overlap.~\cite{bastard,pawel} Therefore, recombination mainly occurs between e-h pairs belonging to the same respective energy levels.~\cite{bastard,pawel}

The recombination of the e-h pair in the second level is therefore possible only from the ``bright" triexciton states.
Like the exciton case, these recombinations give rise to two cross-rectilinearly polarized spectral lines which leave a ground state biexciton in the QD ($XX^{0}$, Fig. \ref{fig:theory}(b)).
The ground state biexciton continues to radiatively decay by a well-studied two-photon radiative cascade,\citep{regelman2001,akopian2006,moreau2001,kiras2002,gammon1996,benson2000,kulakovskii1999} 
potentially providing a source of pairs of entangled photons on demand.~\citep{benson2000,akopian2006,young2006,hafenbrak2007,muller2014} A comprehensive review of two photon radiative cascades in QDs is available in Ref.\cite{poem2012} In the case of this second-level recombination (not studied here), the triexciton forms a direct three-photon radiative cascade.

However, both the ``dark" and the ``bright" triexciton states can recombine radiatively by annihilating a ground level e-h pair. For a thermally-populated ground triexciton state, these recombinations occur in about 5 out of 6 recombinations. In these cases, an excited biexciton is left in the QD, containing one e-h pair in their respective ground levels and one in their respective second energy levels. The excited e-h pair usually relaxes to their ground level by phonon emission~\cite{poemprb2010,kodriano2010,benny2014} before radiative recombination occurs. Thus an indirect photon radiative cascade is formed, involving a phonon emission in addition to the three photons.

There are $2^{4}=16$ possible spin configurations
for the remaining carriers in the excited biexciton. These can be conveniently sorted in the following way, which takes into account same-carrier exchange interactions:
1 e-singlet/h-singlet state, 3 e-singlet/h-triplet
states, 3 e-triplet/h-singlet states, and 9 e-triplet/h-triplet states. All these states were observed in two laser PLE spectroscopy.\citep{benny2011prb} Benny {\it et al.}~\citep{benny2011prb} used the first laser pulse to generate an exciton in the QD and the second delayed, tunable pulse to search for excited biexciton absorption resonances. We used a similar technique, albeit with three laser pulses, to detect the direct ``bright" triexciton optical transitions described in Fig. \ref{fig:theory}(b).

The optical transitions from the four triexciton states to the nine e-triplet/h-triplet excited biexciton states (Fig. \ref{fig:theory}(c)) are observed and identified here using polarization-sensitive PL spectroscopy.
Transitions to biexciton states which involve same-carrier singlet configurations are more difficult to observe in PL, since they rapidly decay non-radiatively to their ground singlet level.
This rapid decay results in spectral broadening of the optical transitions to these states, thereby rendering their identification in PL spectroscopy quite challenging.

Fig. \ref{fig:theory}(b) schematically illustrates the allowed optical transitions
from the triexciton states to the ground biexciton state and, in \ref{fig:theory}(c), the optical transitions to the
nine e-triplet/h-triplet states of the excited biexciton states.
Horizontal (H) (vertical, V) polarized emission
is indicated by blue (red) arrows. The total spin projection of the
initial (final) state is presented to the right of the energy level.
For example, $|+2\rangle$ corresponds to the normalized spin wavefunction
of $|-1_{e}\rangle|+3_{h}\rangle$ where the electrons are in a spin-parallel
down triplet state and the heavy holes are in a spin-parallel up
triplet state. There are 2 allowed transitions to the ground biexciton state and 10 allowed transitions to the excited triplet-triplet states. Taking into account the very small splitting between the ``dark" triexciton
states, the latter 10 transitions yield three almost unpolarized transitions initiating from the ``bright" triexciton states and two strongly rectilinearly polarized  transitions initiating from the ``dark" triexciton states. We note here the one-to-one correspondence between these triexciton-excited biexciton transitions and the exciton-excited biexciton transitions obtained in Ref~\citep{benny2011prb}.
The calculated PL emission spectrum, from the many-carrier configuration-interaction (CI) model discussed in Ref~\citep{benny2011prb}, is provided above the transition scheme. This model takes into account the direct and exchange Coulomb interactions between the quantum-confined carriers. We then use the dipole approximation to calculate the optical transitions between states for a given light polarization.\cite{poem2007} 

In general, the nine e-triplet/h-triplet excited biexciton states are spin blockaded from relaxation to the ground e-singlet/h-singlet biexciton state. There is, however, a relatively efficient spin flip-flop mechanism which permits this relaxation.~\citep{benny2014} In this process, an electron and hole flip their spin mutually due to the enhanced effect of the e-h exchange interaction in the presence of near resonant e-LO phonon Fr\"{o}lich interaction.~\citep{benny2014} The even (0 and $\pm 2$) total spin excited biexciton states efficiently relax this way to the ground state biexciton ($XX^0$), while the odd states ($\pm 1$ and $\pm 3 $) relax to the spin blockaded biexciton $XX^{0}_{T\pm 3}$ states.~\citep{poem2010,kodriano2010}
The relaxation to the $XX^{0}_{T\pm 3}$, proceeds by emission of another photon and leaves the QD with a dark exciton.~\citep{poem2010,schwartz2014} The details of this two-photon radiative cascade are left for a forthcoming publication.

\begin{figure}
\includegraphics[width=1\columnwidth]{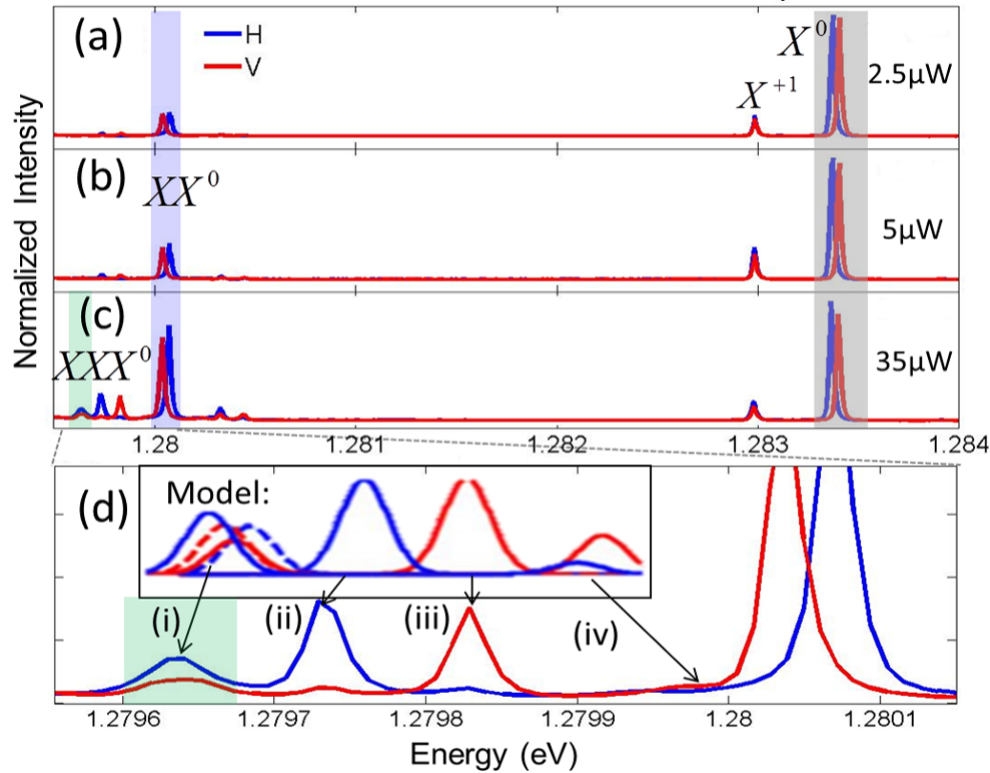}
\protect\caption{Polarization-sensitive PL measurements for a single quantum dot under
varying off-resonant excitation intensities (a-c). Initial state for
recombination is indicated above each emission line. (d) The four
triexcitonic emission lines compared with the calculated PL spectrum
(inset). Roman numerals label emission lines corresponding to the transitions in Fig. \ref{fig:theory}(c). \label{fig:PL}}
\end{figure}

The sample used here was grown by molecular beam epitaxy (MBE)
on a {[}001{]}-oriented GaAs substrate. One layer of self-assembled
InGaAs QDs was deposited in the center of a one-wavelength planar microcavity designed
for enhancing light harvesting resulting from emission due ground level e-h recombination.
As a result, however,  detection of emission due to recombination of e-h
pairs from the second level was practically eliminated.
For optical measurements, the sample was placed inside a cryogenic
tube, maintaining the sample temperature at 4K. A $\times$60, 0.85
numerical aperture \emph{in situ} microscope objective was used to
focus the exciting laser lights on the sample surface and to collect the
emitted PL light. For resonant pulse excitation three synchronously pumped energy-tunable dye lasers of
8 ps pulse width each and 76 MHz repetition rate were used.
For non-resonant continuous wave (cw) excitation, a 445nm diode laser was used.
The collected PL emission was split by a non-polarizing beam splitter, and the polarization of the emitted light in each beam was analyzed using two computer-controlled
liquid-crystal variable retarders and a polarizing beam splitter. The setup thus provided 4 PL collection channels.
In each channel, the light was dispersed by a monochromator and detected by either
a CCD camera or by a single-photon silicon avalanche photodetector.
The setup provided spectral resolution of about 15 $\mu$eV and temporal resolution of about 400 ps.

Fig. \ref{fig:PL} presents polarization-sensitive PL spectra of the
QD at increasing powers of the 445nm light. At low excitation
power, the dominant emission line is that corresponding to
recombination from single excitons. Increasing excitation
power results in increasing biexciton (Fig. \ref{fig:PL}(b)) and ultimately
triexciton (Fig. \ref{fig:PL}(c)) PL emission.
Fig. \ref{fig:PL}(d) presents the measured triexciton emission
next to the calculated spectrum.~\citep{benny2011prb} Good correspondence is observed between
the polarization sensitive measurements and the calculated spectrum providing identification
of the observed PL lines. The three shaded emission lines are the lines used for
the third-order intensity correlation ($g^{(3)}$) measurements of the three-photon radiative cascade of the triexciton.

Potential use of the triexciton and its radiative cascade requires its deterministic generation
in a well-defined spin configuration. Benny {\it et al.}~\cite{benny2011prl}
demonstrated a one-to-one correspondence between the polarization of a resonant laser pulse and the spin of the photogenerated bright exciton. Here, we use a sequence of three laser pulses to demonstrate the same ability for the bright triexciton. The first two $\pi$-pulses deterministically generate the ground state biexciton $XX^0$. A third, slightly delayed $\pi$-pulse, resonantly tuned to the direct ground state biexciton-bright triexciton transition (Fig. \ref{fig:theory}(b)) deterministically generates the triexciton. 

\begin{figure}
\includegraphics[width=1\columnwidth]{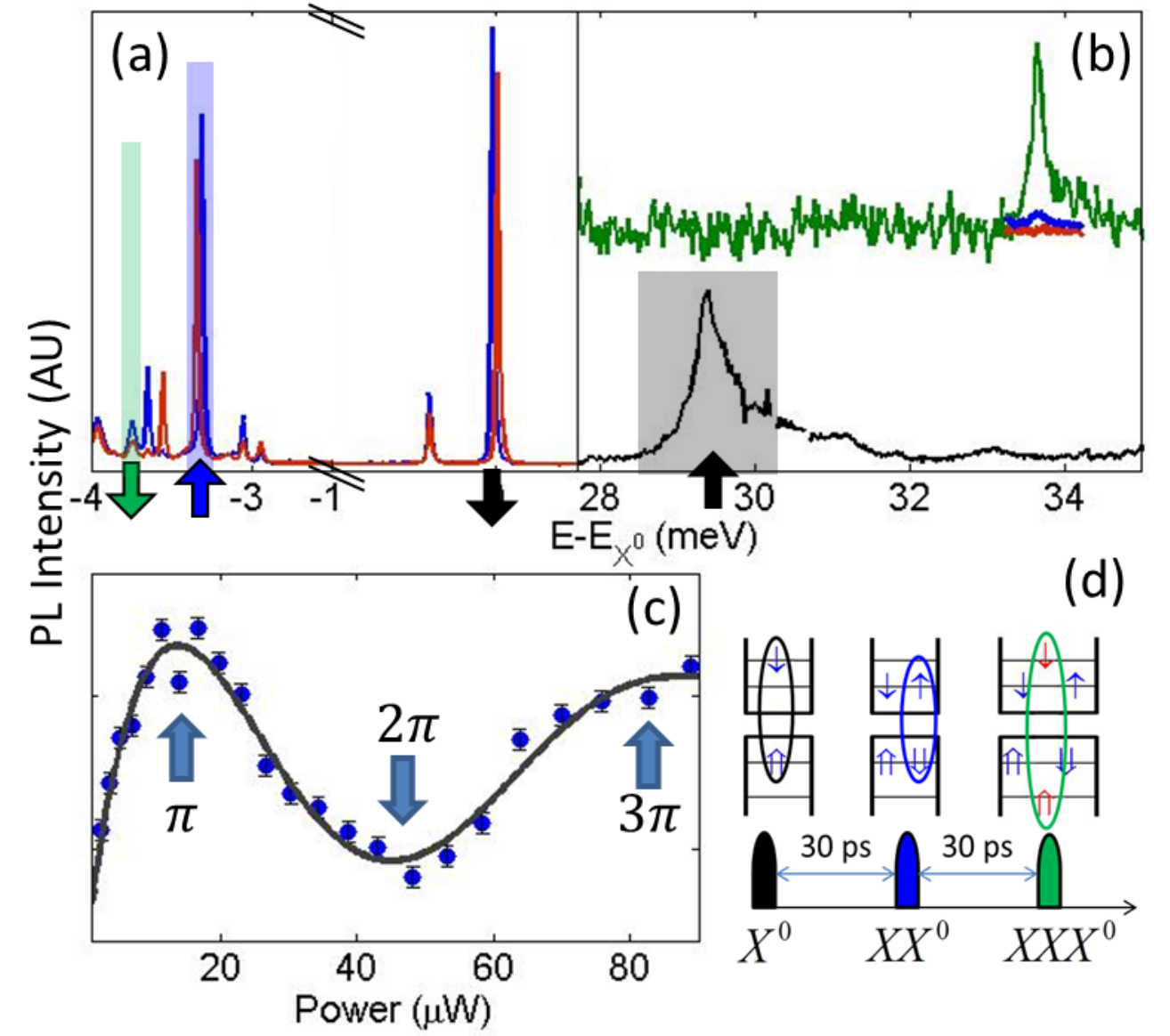}
\protect\caption{(a) Rectilinear polarization-sensitive PL spectra from a strongly excited QD. (b) Black (green) curve is the PLE spectrum of the $X^0$ ($XXX^{0}$) spectral line marked by a downward  black (green) arrow in (a). The first (second) laser pulse is tuned to the $X^{0}$ ($XX^{0}$) absorption resonance indicated by the upward black (blue) arrow. The red (blue) curve is the PLE measurement in the absence of the first (second) pulse (c) Measured (marks) $XXX^{0}$ PL intensity as a function of the average power of the third pulsed laser tuned to the resonance found in (b).  The solid line describes a theoretical fit to the expected Rabi oscillation behavior. (d) Schematic description of the pulse sequence.
 \label{fig:PLE}}
\end{figure}

In  Fig. \ref{fig:PLE}(b), we present three-laser
photoluminescence excitation (PLE) measurements of one of the ``bright"
$XXX^{0}$ PL emission lines, indicated by a green arrow in the PL spectrum
in Fig. \ref{fig:PLE}(a) and corresponding to the emission line labeled
(i) in Fig. \ref{fig:PL}(d). We used 30 ps inter-pulse temporal spacing between the three pulses, as shown in the schematic in Fig. \ref{fig:PLE}(d).  The first pulse is resonantly tuned to an excitonic absorption resonance
at $\sim29$ meV above the $X^{0}$ emission energy, corresponding
to generation of a $p-$level electron and $s-$level hole.\citep{benny2011prb}
The electron relaxes quickly ($\sim7$ ps) to the $X^{0}$ state in
a spin-preserving process.\citep{benny2011prb,benny2011prl} The second
pulse is resonantly tuned to the $X^0$ - $XX^{0}$ transition energy. In Fig. \ref{fig:PLE}(b), the solid green line shows the PL intensity from the $XXX^{0}$ emission line as a function of the
energy of the third pulse. An absorption resonance is clearly visible $\sim34$ meV above the exciton energy, corresponding to the addition of a $p$-level
e-h pair to the $XX^{0}$ and the formation of $XXX^{0}$ as described in Fig. 1(b).
The solid red (blue) line in  Fig. \ref{fig:PLE}(b)
describes the same PLE measurement as that described by the green solid line but without the first (second) laser pulse, verifying that the triexciton PL indeed results only by the three pulses together. In Fig. \ref{fig:PLE}(c), we present the $XXX^{0}$ PL intensity as a function of the average power of the third resonantly tuned pulse. Rabi oscillations are clearly visible, demonstrating that our three-pulse sequence deterministically photogenerates the triexciton. The polarization of the final pulse in the sequence determines the spin configuration of the unpaired $p$-level e-h pair.\citep{benny2011prl} Thus, the $XXX^{0}$ can be deterministically generated in any {\it a priori} well-defined spin configuration.

Intensity correlation measurements are the most common measurement technique to establish the quantum nature of light emitted from single photon sources, such as semiconductor QDs~\cite{dekel1999, michler2000} and nanocrystals,~\cite{michler2000b} and to characterize radiative cascades in QDs. In these cases, $g^{(2)}_{1,2}(\tau)=\langle I_1(t)I_2(t+\tau)\rangle/\left(\langle I_1(t)\rangle \langle I_2(t)\rangle \right)$ is measured using a two-channel Hanbury Brown-Twiss (HBT) apperatus.\citep{HBT,regelman2001,moreau2001,poem2012} Here, $I_{i}(t)$ is the intensity of light at time $t$ on the $i$th detector, $\tau$ is the time between the detection of a photon in detector 1 and detection of a subsequent photon in detector 2 and $\langle\,\rangle$ means temporal average. A radiative cascade is characterized by an asymmetric correlation function, due to the temporal order of the emitted photons. Following the detection of the second photon in a cascade, no detection of emission of the first photon is possible. Therefore ``antibunching"  ($g^{(2)}_{1,2}(\tau)< 1$) is anticipated. However, following the detection of the first photon the probability of detecting the second photon is higher than the steady state probability~\citep{regelman2001,dekel1999} and bunching ($g^{(2)}_{1,2}(\tau)> 1$) is anticipated.~\cite{regelman2001,moreau2001,poemprb2010,kodriano2010,poem2012}

Here, for characterizing three-photon radiative cascades, we use a three-channel HBT apparatus
for measuring the third-order intensity correlation function,\citep{glauber1963,assman2009,elvira2011, stevens2014, rundquist2014}
\begin{equation}
g^{(3)}_{1,2,3}(\tau_{1},\tau_{2})=\frac{\left\langle I_1(t)I_2(t+\tau_1)I_3(t+\tau_2)\right\rangle }{\left\langle I_1(t)\right\rangle \left\langle I_2(t)\right\rangle \left\langle I_3(t)\right\rangle }
\end{equation}
 where $\tau_{1}$, ($\tau_{2}$) is the time between the detection of the first photon by the first detector and the time of detecting the second (third) photon by the second (third) detector.
 Using these definitions it is straightforward to show that:
 \begin{align}
\langle g^{(3)}_{1,2,3}(\tau_{1},\tau_{2})\rangle_{\tau_{2}}=g^{(2)}_{1,2}(\tau_{1})\nonumber \\ \langle g^{(3)}_{1,2,3}(\tau_{1},\tau_{2})\rangle_{\tau_{1}}=g^{(2)}_{1,3}(\tau_{2})
\end{align}

\begin{figure}
\includegraphics[width=1\columnwidth]{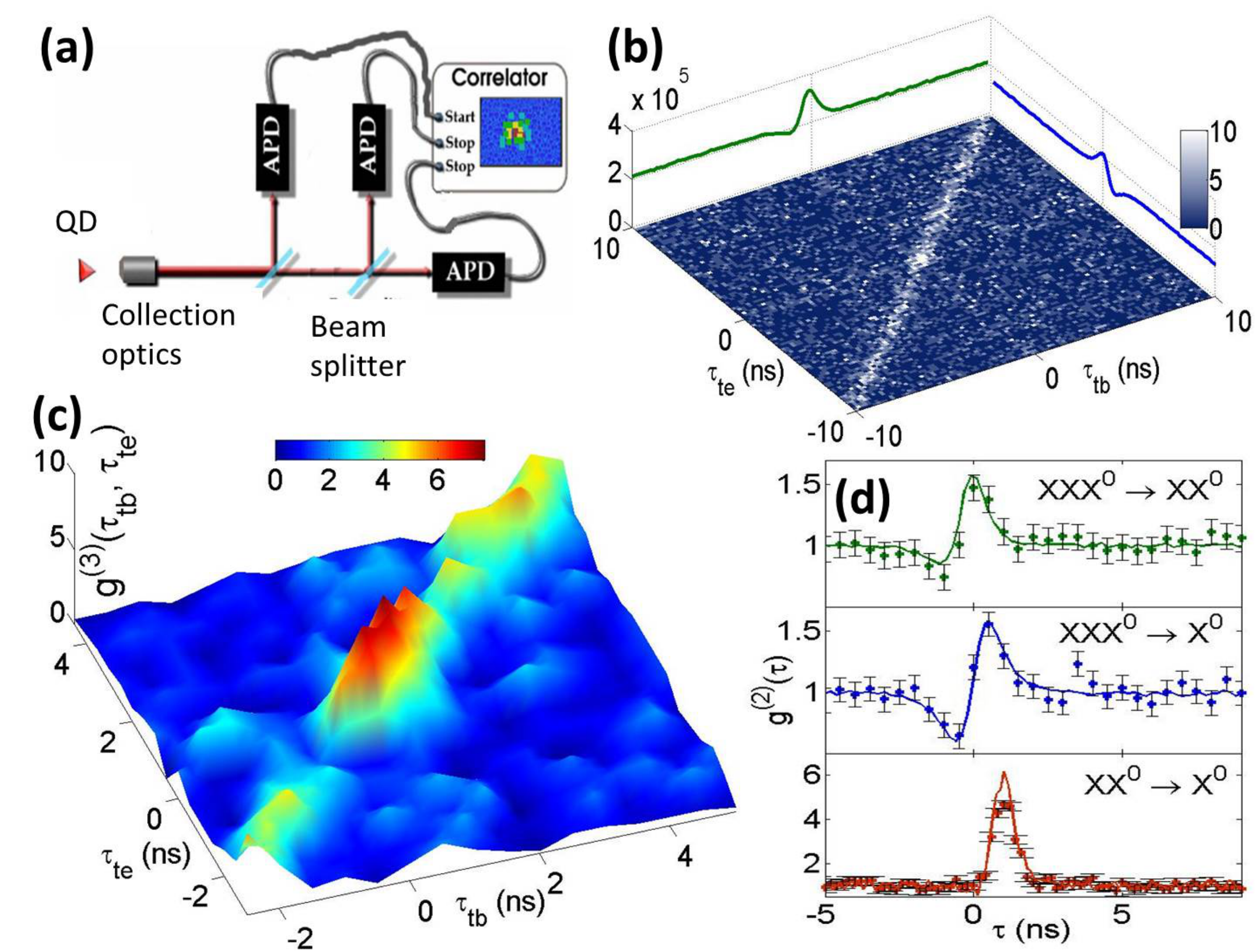}

\protect\caption{(a) Schematic description of the detection setup. (b) Histogram displaying the number of two-photon (edges) and three-photon (center) events as a function of  $\tau_{tb}$ and $\tau_{te}$. 
(c) 2D normalized histogram displaying the measured $g^{(3)}_{1,2,3}(\tau_{tb},\tau_{te})$, as deduced from (b). (d) The solid curves represent $g_{1,2}^{(2)}(\tau_{tb})$, $g_{1,3}^{(2)}(\tau_{te})$ and $g_{2,3}^{(2)}(\tau_{be})$ extracted from the corresponding two-photon events. Overlaid marks with error-bars represent second order intensity correlation functions extracted from the third order intensity correlation function by temporal averaging (Eq. 2). \label{fig:g3}}
\end{figure}

Fig. \ref{fig:g3}(a) schematically describes the three-channel HBT
system used to measure the third order intensity correlation function.
Detection times were recorded by a four-channel PicoQuant HydraHarp
single-photon event timer.\citep{wahl2007} Time differences between detection events
on every channel were deduced and used to generate a multi-dimensional histogram.
This way, three second order ($g^{(2)}_{1,2}(\tau_{tb})$, $g^{(2)}_{1,3}(\tau_{te})$, $g^{(2)}_{2,3}(\tau_{be})$) where $\tau_{tb}$, $\tau_{te}$ and $\tau_{be}=\tau_{te}-\tau_{tb}$ are the triexciton-biexciton, triexciton-exciton and biexciton-exciton time differences respectively) and one third order ($g^{(3)}_{1,2,3}(\tau_{tb},\tau_{te})$) intensity correlation measurements were simultaneously carried out.

Fig. \ref{fig:g3} presents the measured intensity correlation functions
for the three-photon radiative cascade initiating from the $XXX^{0}$.
The first detector was tuned to the spectral line corresponding
to $|+1\rangle\pm|-1\rangle\rightarrow|2\rangle\pm|-2\rangle$ transition
from $XXX^{0}$, the PL emission line labeled (i) in Fig. \ref{fig:PL}(d)
and used for the PLE measurements in Fig. \ref{fig:PLE}(b). The second and third
detectors were tuned to the $XX^{0}$ emission line (blue
shading in Fig. \ref{fig:PL}(a-c)) and to the $X^{0}$ emission line
(black shading in Fig. \ref{fig:PL}(a-c)), respectively. Fig. \ref{fig:g3}(b)
shows a two dimensional histogram displaying the number of three photon events as a function of  $\tau_{tb}$ and $\tau_{te})$. The number of measured two photon events in which only triexciton and biexciton (exciton) photons were detected are plotted as a function of $\tau_{tb}$ ($\tau_{te}$) to the right (above) the 2D histogram in Fig. \ref{fig:g3}(b). We note the almost three orders of magnitude larger statistics accumulated for two photon events than for three photon events. Specifically, of the $\sim 3.4 \times 10^7$ recorded events, $\sim$ 99.95\% are two-photon events. There are a total of $\sim$16,000 recorded three-photon events in this data set. By comparing the number of recorded three-photon events with that of the two-photon events, one can directly obtain the light harvesting efficiency of the experimental setup. Assuming that all two-photon events result from triexciton radiative cascades, then the photon collection efficiency $\eta_i$ of channel $i$ is given by $\eta_i=N_{123}/N_{jk}$
where $N_{123}$ ($N_{jk}$) is the total number of three (two) photon events (recorded in detectors $j$ and $k$). 
The efficiencies of the three PL collection channels in our setup ranging from 1 in 600 to 1 in 1000.

In Fig. \ref{fig:g3}(c), we present the measured third order intensity correlation function as obtained by normalizing the 2D histogram of Fig. \ref{fig:g3}(b).
In Fig. \ref{fig:g3}(d), we present by solid lines three measured second order intensity correlation function as obtained by normalizing the large statistics 1D histograms (from two photons events) of Fig. \ref{fig:g3}(b).
As a validity check of our approach, we obtained the second order intensity correlation functions also by temporal summation over the measured third order correlation function (Eq. 2). These data points are overlaid on the much higher statistics curves obtained from the measured two photon events and the agreement is very good.
Clearly, with higher statistics (e.g. by improving the light collection efficiency) these measurements could be applied for studying dynamical effects and coherence loss mechanisms which occur during the radiative cascades.

In conclusion, we have demonstrated deterministic generation of the QD-confined triexciton via a three laser pulse sequence. We also identified and characterized the PL emission lines due to
all possible e-h recombinations from the ground states of the triexciton. Particularly, we investigated and conclusively characterized an indirect three photon radiative cascade initiating from the triexciton using novel third-order intensity cross-correlation measurements. The experimental tools developed for characterization of multiphoton radiative cascades and higher-order excitonic states in semiconductor QDs are essential for further understanding many-carrier states in QDs and for novel protocols which require higher order quantum correlations between carriers confined in these QDs and photons that they emit.

\begin{acknowledgments}
The support of the Israeli Science Foundation (ISF),
the Technion's RBNI and the Israeli Nanothecnology Focal Technology Area on ``Nanophotonics for Detection" are gratefully acknowledged.
\end{acknowledgments}

\bibliographystyle{unsrtnat}
\bibliography{triexciton_bib}

\end{document}